\documentstyle[prd,aps,epsfig,floats]{revtex}
\begin{document}

\draft
\renewcommand{\topfraction}{0.8} 
\newcommand{\beq}{\begin{equation}}
\newcommand{\eeq}{\end{equation}}
\newcommand{\bea}{\begin{eqnarray}}
\newcommand{\eea}{\end{eqnarray}}
\newcommand{\pbar}{\not{\!\partial}}
\newcommand{\dbar}{\not{\!{\!D}}}
\def\lsim{\:\raisebox{-0.75ex}{$\stackrel{\textstyle<}{\sim}$}\:}
\def\gsim{\:\raisebox{-0.75ex}{$\stackrel{\textstyle>}{\sim}$}\:}
\twocolumn[\hsize\textwidth\columnwidth\hsize\csname 
@twocolumnfalse\endcsname

\title{Leptogenesis as the source of gravitino dark matter and density
  perturbations}   
\author{Rouzbeh Allahverdi$^{a}$ and Manuel Drees$^{b}$}
\address{$^{a}$Theory Group, TRIUMF, 4004 Wesbrook Mall, Vancouver, British 
Columbia, V6T 2A3, Canada.\\
$^{b}$Physikalisches Institut der Universit\"at Bonn, Nussallee 12, D-53115, 
Bonn, Germany.}
\date{\today} 
\maketitle

\begin{abstract}
  
  We investigate the possibility that the entropy producing decay of a
  right-handed sneutrino condensate can simultaneously be the source of the
  baryon asymmetry, of gravitino dark matter, and of cosmological density
  perturbations. For generic values of soft supersymmetry breaking terms in
  the visible sector of $1-10$ TeV, condensate decay can yield the dark matter
  abundance for gravitinos in the mass range 1 MeV to 1 TeV, provided that the
  resulting reheat temperature is below $10^6$ GeV. The abundance of thermally
  produced gravitinos before and after sneutrino decay is then negligible. We
  consider different leptogenesis mechanisms to generate a sufficient
  asymmetry, and find that low-scale soft leptogenesis works most naturally at
  such temperatures. The condensate can easily generate sufficient density
  perturbations if its initial amplitude is $\sim {\cal O}(M_{\rm GUT})$, for
  a Hubble expansion rate during inflation $> 10^9$ GeV. Right-handed
  sneutrinos may therefore at the same time provide a source for baryogenesis,
  dark matter and the seed of structure formation.

\end{abstract}

\vskip2pc]

%%%%%%%%%%%%%%%%%%%%%%%%%%%%%%%%%%%%%%%%%%%%%%%%%%%%%%%%%%%%%%%%%%%%%%%%%%%
\section{Introduction}
\setcounter{footnote}{0}

The recent WMAP measurement of the temperature anisotropy of the cosmic
microwave background (CMB) has led to the precise determination of many
cosmological parameters~\cite{wmap}. Among these, the density of baryons
$\Omega_{\rm b}$ and total matter density $\Omega_{\rm m}$ are of particular
interest and importance ($\Omega_X$ denotes the energy density of species $X$
normalized to the critical energy density of the universe). For the WMAP data
only, the best fit parameters are $\Omega_{\rm b} h^2 = 0.024 \pm 0.001$ and
$\Omega_{\rm m} h^2 = 0.14 \pm 0.02$, where the present value of the Hubble
parameter $h = 0.72 \pm 0.05$ (in units of 100 km sec$^{-1}$ Mpc$^{-1}$)
\cite{wmap}. The inferred value of $\Omega_{\rm b}$ implies that we live in a
baryon asymmetric universe where $\eta_{\rm B} \equiv (n_{\rm B}-n_{\bar{\rm
    B}})/s \simeq 0.9 \times 10^{-10}$, $s$ being the entropy density of the
universe. This is in good agreement with an independent determination from Big
Bang nucleosynthesis (BBN)~\cite{bbn}. The determined value of $\Omega_{\rm
  m}$, on the other hand, implies that most of the matter in universe is
non-baryonic and dark. Since a period of inflation~\cite{infl} washes away any
existing baryon and matter density, it seems inevitable that both baryons and
dark matter should be generated in the post-inflationary epoch. This has been
the focus of intensive research activities in both particle physics and
cosmology for the past two decades.

The production of the baryon asymmetry from a baryon symmetric universe
requires that three conditions are met: $B$ violation, $C$ and $CP$ violation,
and departure from thermal equilibrium~\cite{sakharov}.  Actually, $B+L$
violating sphaleron transitions, which are active at temperatures $100~{\rm
  GeV} \lsim T \leq 10^{12}$ GeV~\cite{krs}, imply that any mechanism
operating at $T > 100$ GeV must generate a $B-L$ asymmetry. The final baryon
asymmetry is then given by $B=a(B-L)$, where $a=28/79$ in case of the standard
model (SM) and $a=8/23$ for the minimal supersymmetric standard model
(MSSM)~\cite{khlebnikov}.

Leptogenesis is an attractive mechanism for producing a $B-L$
asymmetry~\cite{fy}. This scheme postulates the existence of right-handed (RH)
neutrinos $N_i$ with a lepton number violating Majorana mass $M_N$; since the
$N_i$ are singlets under the SM gauge group, $M_N$ can be
much larger than the electroweak scale. This provides an elegant explanation
for the small masses of the light neutrinos via the see-saw
mechanism~\cite{seesaw}. Moreover, a lepton asymmetry can be generated from
the out-of-equilibrium decay of the RH neutrinos, provided $CP$ violating
phases exist in the neutrino Yukawa couplings; this lepton asymmetry will be
partially converted to a baryon asymmetry via sphaleron
effects~\cite{fy,luty,plumacher}.

The RH neutrinos can be produced thermally or non-thermally in the early
universe. In the thermal scenario~\cite{plumacher}, the generation of an
acceptable lepton asymmetry requires the mass $M_1$ of the lightest RH
neutrino and the temperature of the thermal bath to exceed
$10^8$~GeV~\cite{buchmuller,sacha,gnrrs} (unless RH neutrinos are
degenerate~\cite{pilaftsis}, or for specific neutrino mass 
models~\cite{strumia}). However, in many supersymmetric theories this
is marginally compatible with the upper bound from thermal gravitino
production~\cite{ellis} on the reheat temperature $T_{\rm R}
\leq(10^{6}-10^{9})$ GeV~\cite{bbnbound}.\footnote{Non-thermal production of
  gravitinos during preheating~\cite{non} does not give rise to severe bounds
  in realistic models of inflation~\cite{abm,nps}.}  Alternatively, RH
neutrinos can be produced through the perturbative~\cite{pert} or
non-perturbative~\cite{gprt,tachyon} decay of the inflaton; in this case the
reheat temperature can be significantly below $M_N$. Non-thermal leptogenesis
can also be achieved without exciting on-shell RH neutrinos~\cite{bb,am,dlr}.

In supersymmetric models one also has the RH sneutrinos which can serve as an
additional source for leptogenesis~\cite{cdo}. They are produced along with
the neutrinos in a thermal bath or during reheating, and with much higher
abundances in preheating~\cite{bdps}.  Moreover, there are two unique
possibilities for leptogenesis from RH sneutrinos. First, they can acquire a
large vacuum expectation value (VEV) if their mass during inflation is less
than the Hubble expansion rate at that epoch $H_{I}$. This condensate starts
oscillating once $H \simeq M_N$, thereby automatically satisfying the
out-of-equilibrium condition. The decay of the sneutrino condensate can then
yield the desired lepton asymmetry in the same fashion as neutrino decay
does~\cite{my,hmy}. The second possibility is to generate a lepton asymmetry
in the RH sneutrino sector which will be transferred to the light (s)leptons
upon its decay. This asymmetry can be produced via sneutrino couplings to the
inflaton~\cite{bmp,adm}, or from soft supersymmetry breaking
effects~\cite{nir1,agr,ad3,chun,goran,nir2}.

As already stated, the WMAP measurement also implies the existence of a large
amount of non--baryonic dark matter (DM) in the universe. Other evidence for
the existence of DM comes from the analysis of galactic rotation curves, from
determinations of the masses of clusters of galaxies (i.e. using the X--ray
temperature or gravitational lensing techniques), and from attempts to model
structure formation in the universe \cite{dmrev}. In supersymmetric models
with exact $R$ parity the most natural particle physics DM candidate is the
lightest superparticle (LSP). Since superparticles are odd under $R$ parity
while ``ordinary'' particles are even, exact $R$ parity implies that the LSP
is stable.

The most widely studied LSP candidate, in particular in the context of dark
matter, is the lightest neutralino $\tilde \chi_1^0$ \cite{jung}. It has
several appealing features. Its interactions with ordinary matter, while
small, might be sufficient to allow for its detection \cite{jung}. Moreover,
the $\tilde \chi_1^0$ mass, as well as the parameters that determine its
couplings, can be measured at colliders \cite{colliders} (once they reach
sufficient energy to produce on-shell superparticles). For some ranges of
these parameters $\tilde \chi_1^0$ has the right thermal relic density. All
these statements also hold in simple, constrained versions of the MSSM, which
allow to describe supersymmetry breaking with only a few free parameters
\cite{msugra}.

However, neutralinos decouple from the thermal bath at a relatively low
temperature $\sim m_{\tilde \chi_1^0}/20$. This makes it rather difficult to
find a common origin for the creation of the baryon asymmetry and of dark
matter. Such a common origin might be hinted at by the observation that
$\Omega_{\rm b}$ and $\Omega_{\rm DM}$ are of similar order of magnitude. Even
ignoring this coincidence, it would certainly be more economical to find a
common mechanism for explaining two seemingly unrelated observations.

One such explanation is based on the late decaying $Q-$ball scenario in models
with gravity-mediated supersymmetry breaking~\cite{fy1} (for a similar
scenario in gauge-mediated models, see~\cite{fy2}). It is known that in many
cases coherent oscillations of flat directions carrying a non-zero baryon
number fragment into non-topological solitons, called supersymmetric
$Q-$balls~\cite{qball}. If the initial VEV of the flat direction is
sufficiently large, $Q-$ball decay (which typically releases three LSPs per
unit of baryon number) can occur below the LSP freeze-out
temperature~\cite{em}. This would give the right DM density for $m_{\tilde
  \chi_1^0} \simeq 2 m_p$, which is well below the current experimental lower
bound in constrained supersymmetric models \cite{msugra}. In this scenario
overproduction of dark matter can only be avoided if the LSP is dominantly a
wino or Higgsino, {\em and} if the $Q-$ball decay temperature is tuned to be
sufficiently high to allow for significant $\tilde \chi_1^0$ annihilation (but
below the temperature where $\tilde \chi_1^0$ is fully thermalized)~\cite{fh}.

The LSP might also reside in the ``hidden'' or ``secluded'' sector, thought to
be responsible for the spontaneous breakdown of supersymmetry. Here the most
widely studied candidate is the gravitino. Its interactions are determined
uniquely by its mass (and the soft breaking parameters in the visible sector).
If produced thermally, the gravitino mass density increases linearly with the
reheat temperature $T_{\rm R}$ and inversely with the gravitino mass $m_{3/2}$
\cite{ellis}. Gravitinos are also produced non-thermally from the decay of the
next-to-lightest superparticle (NLSP), typically the lightest neutralino or a
scalar lepton. One then has to require that these decays do not violate
constraints from BBN \cite{bbn-gravi}. This constraint is easier to satisfy if
the NLSP is not a bino. By adding thermal and non--thermal gravitino
production, the gravitino can have the required relic density over quite wide
ranges of the parameter space \cite{gravi-dm}.  It is even possible to find
some combinations of $T_{\rm R}$ and $m_{3/2}$ that give successful thermal
leptogenesis and thermal gravitino dark matter \cite{bbp}. However, in this
scenario the connection between $\Omega_{\rm b}$ and $\Omega_{\rm DM}$ is very
indirect.

Recently, it has been shown~\cite{hmy,iy} that the decay of a RH sneutrino
condensate which dominates the energy density of the universe can result in
successful leptogenesis as well as dark matter production for gravitino masses
in the MeV range. Such small gravitino masses are expected in models with
gauge mediated supersymmetry breaking \cite{gmsb}, but can also occur in
models with gravity mediated supersymmetry breaking for certain
(non-canonical) choices of the K\"ahler metric \cite{light-gravi}. In this
scenario the late decay of the lightest sneutrino, while generating the lepton
asymmetry, reheats the universe and gives rise to thermal production of
gravitino dark matter. Obtaining the correct abundances of baryon asymmetry
and and dark matter then determine the reheat temperature of the universe
$T_{\rm R} \simeq 10^6$ GeV and the gravitino mass $m_{3/2} \simeq 10$ MeV,
respectively~\cite{iy}. 

In this note we explore the prospect of scenarios offering an even closer
connection between the baryon and gravitino densities, where both are produced
{\em directly} from (s)neutrino decay. The gravitino channel will be
kinematically open if supersymmetry breaking results in a mass difference
between the RH sneutrino and neutrino which is $> m_{3/2}$. This mass
splitting is essentially determined by the bilinear soft breaking parameter
$B$ associated with the (supersymmetric) RH neutrino Majorana mass term.  We
find that gravitino dark matter with mass $10~{\rm keV} \lsim m_{3/2} \lsim
10$ TeV can be produced from sneutrino decay for $B$ in the range $1-10$ TeV,
compatible with the expected size of visible sector soft breaking terms,
provided that the resulting reheat temperature $T_{\rm R} < 10^6$ GeV.
Additional considerations reduce this range to 1 MeV $\lsim m_{3/2} \lsim$ 1
TeV. As we will see, among different leptogenesis mechanisms, low-scale soft
leptogenesis is preferred to generate the observed asymmetry at such
relatively low values of $T_{\rm R}$. Sufficient dilution of the gravitinos
which are thermally produced at earlier stages typically requires that the
initial VEV of the sneutrino condensate be $\simeq 10^{16}-10^{17}$ GeV.
Fluctuations in the sneutrino energy density can then generate cosmological
density perturbations via the curvaton mechanism~\cite{lyth,mt} if the Hubble
expansion rate during inflation is $\sim 10^{11}$ GeV. A RH sneutrino
condensate may therefore address three major cosmological issues
simultaneously.

The rest of this paper is organized as follows. In Section II we will discuss 
gravitino production in sparticle decays, specializing to the case for a RH 
sneutrino. We will turn to leptogenesis from a sneutrino condensate in 
Section III. In Section IV we will address generation of cosmological density 
perturbations from a sneutrino condensate. Additional issues such as effects 
of inflaton decay on the condensate dynamics and vice versa will be 
considered in Section V. We will conclude by summarizing our results in the 
closing Section VI.                                       
               
%%%%%%%%%%%%%%%%%%%%%%%%%%%%%%%%%%%%%%%%%%%%%%%%%%%%%%%%%%%%%%%%%%%%%%%%%%%%%%
%%%%%%%%%%%%%%%%%%%%%%%%%%%%%%%%%%%%%%%%%%%%%%%%%%%%%%%%%%%%%%%%%%%%%%%%%%%%%%
\section{Gravitino production from sneutrino decay}
\setcounter{footnote}{0} We work in the framework of the MSSM~\cite{nilles} 
augmented with three RH
neutrino multiplets in order to accommodate neutrino masses via the see-saw
mechanism~\cite{seesaw}. The relevant part of the superpotential is
\beq \label{superpot}
W \supset {1 \over 2} M_N {\bf N} {\bf N} + {\bf h} {\bf H}_u {\bf N} {\bf L},
\eeq
where ${\bf N}$, ${\bf H}_u$, and ${\bf L}$ are supermultiplets containing the
RH neutrinos $N$ and sneutrinos $\tilde N$, the Higgs field giving mass to the
top quark and its superpartner, and the left--handed (s)lepton doublets,
respectively. $h$ are the neutrino Yukawa couplings; for simplicity, family
indices on $M_N$, $h$, ${\bf N}$, and ${\bf L}$ are omitted. We work in the
basis where the Majorana mass matrix is diagonal.

In addition to the supersymmetry conserving part of the scalar potential for 
$\tilde N$, one also has soft terms from 
low-energy supersymmetry breaking~\cite{nilles}
\beq \label{low}
V_{\rm soft} \supset m^2_0 \vert {\tilde N} \vert^2 + (B M_N {\tilde N}^2 + A
h {\tilde N} H_u  {\tilde L} + {\rm h.c.}),
\eeq
where $m_0$ and $\vert B \vert$ typically are ${\cal O}(1~{\rm TeV})$.
Supersymmetry breaking by the energy density of the universe also makes
contributions $\propto H$ to the soft terms~\cite{drt1}. However, these will
be negligible, since $H \ll 1$ TeV during the relevant epoch when the lightest
RH (s)neutrino decays, as we will see.

In models of local supersymmetry, the gravitino appears as spin-$3/2$ partner
of the graviton. It acquires a mass $m_{3/2}$ after supersymmetry breaking
through the superHiggs mechanism~\cite{nilles}. Gravitino couplings to other
particles are suppressed by inverse powers of the reduced Planck mass $M_{\rm
  P} \simeq 2.4 \cdot 10^{18}$ GeV, and so is the rate for its decay to or
production from matter and gauge fields and their superpartners.  Depending on
the model, the gravitino mass could be as small as 1 meV or as large as tens
of TeV. Not surprisingly, the role the gravitino plays in cosmology depends on
the gravitino mass.

If the gravitino is not the LSP, it will decay to particle-sparticle pairs at
a rate $\sim m^3_{3/2}/M^2_{\rm P}$. In the early universe, gravitinos are
produced mainly via scatterings of gauge and gaugino quanta in the thermal
bath.  If $m_{3/2}$ is of the order of the gaugino masses, the helicity $\pm
3/2$ and helicity $\pm 1/2$ states will be produced at approximately the same
rate and we will have~\cite{ellis}
\beq \label{nlspgrav}
{n_{3/2} \over s} \simeq 10^{-12} \left({T_{\rm R} \over 
10^9~{\rm GeV}}\right).
\eeq
For $m_{3/2} \simeq 100~{\rm GeV}-1$ TeV, most gravitinos decay after the
onset of BBN and can distort the abundance of light elements synthesized in
that epoch. The BBN constraints on the abundance of gravitinos then lead to
the bound $T_{\rm R} \leq (10^{6}-10^{9})$ GeV~\cite{bbnbound}. On the other
hand, if $m_{3/2} > 20$ TeV, as in models with anomaly mediated supersymmetry
breaking \cite{anomaly}, gravitinos decay sufficiently early to evade the BBN
constraint. However, LSPs produced in such decays should not overclose the
universe. This will result in an upper bound on $T_{\rm R}$ if gravitino decay
occurs below the LSP freeze-out temperature~\cite{ad2}. This bound can be
relaxed in those parts of the parameter space which allow a more efficient LSP
annihilation and, in consequence, a lower freeze-out temperature~\cite{fh}.

If the gravitino is the LSP, it will be stable, and hence a dark matter
candidate, provided that $R$-parity is unbroken. Then, assuming that gaugino
masses are $\gg m_{3/2}$ in this case, the helicity $\pm 1/2$ states are
dominantly produced and the fractional energy density of the gravitino will be
given by~\cite{murayama1,murayama2}
\beq \label{lspgrav}
\Omega_{3/2} h^2 \simeq {\left({M_{\tilde g} \over 1~{\rm TeV}}\right)}^2 
\left({10~{\rm MeV} \over m_{3/2}}\right) \left({T_{\rm R} \over 
10^6~{\rm GeV}}\right),
\eeq
where $M_{\tilde g}$ is the gluino mass. Then, for $M_{\tilde g} \sim 500$
GeV, the overclosure bound leads to the constraint $T_{\rm R} \leq 10^8
m_{3/2}$.

Note that any scenario which attempts to directly link dark matter to
leptogenesis can only work if the gravitino is the LSP, or if neutralino LSPs
are produced from gravitino decay. This is because leptogenesis can only work
at temperature $T \gsim 100$ GeV where sphalerons are still active. At these
high temperatures neutralinos are still in thermal equilibrium, so their
ultimate relic density will be independent of the number of neutralinos
produced from (s)neutrino decay. Here we will consider models with a
gravitino LSP where such a link exists. We will treat $m_{3/2}$ and the soft
breaking terms in the visible sector, in particular $B$, as independent free
parameters, i.e. we will not assume any specific form for the K\"ahler metric
which determines the relative size of these terms \cite{nilles}. 

We want (s)neutrino decays to be the dominant source of relic gravitinos. When
the difference between sparticle ${\tilde X}$ and particle $X$ masses $\vert
m_{\tilde X} - m_X \vert \gg m_{3/2}$, which may well be the case if the
gravitino is the LSP, helicity $\pm 1/2$ gravitinos are mainly produced in
sparticle decays. These states essentially interact like the Goldstinos; the
relevant couplings are~\cite{murayama1,murayama2}
\beq \label{effective}
{\cal L} \supset {m^2_{\tilde X} - m^2_X \over \sqrt{3} m_{3/2} M_{\rm P}} 
{\tilde X}^* {\bar \psi} \left({1 + \gamma_5 \over 2}\right) X + {\rm h.c.},
\eeq
leading to the partial sfermion decay width
\beq \label{partial}
\Gamma_{{\tilde X} \rightarrow X + \psi} \simeq {1 \over 48 \pi} 
{(m^2_{\tilde X} - m^2_X)^4 \over 
m^2_{3/2} M^2_{\rm P} m^3_{\tilde X}}\, .
\eeq
Note that a smaller $m_{3/2}$ leads to a more efficient gravitino production.

Here we focus on the production of gravitinos from the decay of the lightest
RH (s)neutrino.\footnote{Related studies have been performed for gravitino
production from the decay of the inflaton~\cite{nop}, the superpartners of
stable neutral relics~\cite{aem}, and moduli~\cite{yamaguchi}.} Since the
heavier RH (s)neutrinos essentially play no role in our analysis, we will omit
the generation index on the (s)neutrino field. The first two terms
in~(\ref{low}) contribute to the mass difference between $\tilde N$ and $N$,
but for $M_N \gg 1$ TeV the $B-$term contribution will be dominant. It will
result in two sneutrino mass eigenstates ${\tilde N}_1$ and ${\tilde N}_2$,
with approximate eigenvalues $M_N - \vert B \vert$ and $M_N + \vert B \vert$
respectively. Note that $B$ can be made real and positive by a phase
transformation of $\tilde N$. In that basis, which we will use from now on,
$\tilde N_2$ and $\tilde N_1$ simply are the real and imaginary parts $\tilde
N_R$ and $\tilde N_I$ of $\tilde N$. Thus, for $B > m_{3/2}$, the
decay channels ${\tilde N}_{\rm R} \rightarrow N + {\rm \psi}$ and $N
\rightarrow {\tilde N}_{\rm I} + {\rm \psi}$ are kinematically open.

The evolution of $\tilde N$ with time has been studied in detail in
ref.\cite{ad3}. Since $\tilde N_R$ and $\tilde N_I$ evolve independently, the
co-moving $\tilde N_R$ number density will remain essentially constant for
Hubble parameter $H > \Gamma_N$, where $\Gamma_N \simeq |h|^2 M_N/(4\pi)$ is
the total $\tilde N_R$ decay width. The relative number densities of $\tilde
N_R$ and $\tilde N_I$ are therefore set by the phase of $\tilde N$ at the end
of inflation. Generically one expects this phase to be ${\cal O}(1)$, and
hence $\tilde N_R$ and $\tilde N_I$ to have comparable densities. Here we
assume that the sneutrinos dominate the energy density of the universe before
they decay. In that case most of today's entropy density originates from
sneutrino decays, so that the effective reheat temperature $T_{\rm R} \simeq
\sqrt{M_{\rm P} \Gamma_N}/2$ (for $g_* \simeq 225$ relativistic degrees of
freedom). Moreover, the entropy density just after $\tilde N$ decay satisfies
$s = 4 \rho / (3 T_{\rm R}) \simeq 4 M_N n_{\tilde N} / (3 T_{\rm R})$, where
$\rho$ is the energy density and $n_{\tilde N}$ is the sneutrino number
density just before their decay. Then, for $B \gg m_{3/2}$,
eq.~(\ref{partial}) leads to
\beq \label{dm}
\left({n_{3/2} \over s}\right)_{\rm decay} \simeq 1.6 \cdot 10^{-2} f_R {B^4
  \over m^2_{3/2} T_{\rm R} M_{\rm P}}, 
\eeq
where we have introduced the quantity $f_R = n_{\tilde N_R} / n_{\tilde N} <
1$, which we expect to be ${\cal O}(1/2)$.

Note that the gravitino abundance from $\tilde N_R$ decay is independent of
$M_N$. Since gravitinos are fermions, their occupation number is limited by
Pauli blocking to be $\leq 1$. In addition, the available phase space in
sneutrino decay constrains the physical momentum of produced gravitinos
$k_{3/2} \leq B $. This implies an upper limit
\beq \label{dmmax} 
{\left({n_{3/2} \over s}\right)}_{\rm max} \simeq 3.2 \cdot
10^{-4} \left({B \over T_{\rm R}}\right)^3 \eeq
on the gravitino abundance. Here we have assumed maximum occupation number
throughout the available phase space. This is a valid approximation despite
the fact that $k_{3/2}$ is narrowly peaked around $B$ at the time of
production. The reason is that the sneutrino decay does not occur instantly.
Gravitinos from early sneutrino decays will be redshifted. In the time scale
of interest, i.e.  ${\tilde N}$ lifetime, $k_{3/2}$ will therefore sweep the
phase space due to the Hubble expansion.\footnote{In eq.(\ref{dmmax}) we have
assumed that gravitinos do not thermalize. This is true for the combinations
of $m_{3/2}$ and $T_{\rm R}$ of relevance to us.} It can be seen
from~(\ref{dm}) and~(\ref{dmmax}) that the gravitino abundance will be
saturated at its maximum value for
\beq \label{satur}
m_{3/2} \leq m^{\rm satur}_{3/2} = 7 \left({f_R B T^2_{\rm R} \over 
M_{\rm P}}\right)^{1/2}.
\eeq
On the other hand, for gravitinos to be dark matter, we need
\beq \label{dmgrav}
\left({n_{3/2} \over s}\right)_{\rm DM} \simeq 3 \times 10^{-10} 
\left({1~{\rm GeV} \over m_{3/2}}\right).
\eeq

In order for sphalerons to be active (which is required for leptogenesis) we
need $T_{\rm R} \geq 200$ GeV. On the other hand, eqs.(\ref{dm}) and
(\ref{dmgrav}) show that sneutrino decays produce the desired amount of
gravitino dark matter if
\beq \label{en1}
m_{3/2} = 0.1 \ {\rm GeV}\cdot f_R \cdot \frac {200 \, {\rm GeV} } {T_{\rm R}}
\cdot \left( \frac {B} {1 \ {\rm TeV} } \right)^4 .
\eeq
This desired value of the gravitino mass will be above the saturation value
(\ref{satur}), if
\beq \label{en2}
B > 105 \ {\rm GeV} \cdot \left( \frac {T_{\rm R}} {200 \ {\rm GeV}} 
\right)^{4/7} \cdot \left( \frac {0.5} {f_R} \right)^{1/7}.
\eeq
For $B = 1$ TeV, this allows reheat temperatures up to 10 TeV. In the
framework of leptogenesis, this is still a relatively low temperature. 
$T_{\rm R} = 10^6$ GeV can only be tolerated for $B > 14$ TeV. Soft breaking 
masses of this magnitude are used for first and second generation sfermions in
so-called inverted hierarchy models \cite{invert}. Much larger values of $B$
do not seem plausible to us. If the constraint (\ref{en2}) is violated, the
gravitino relic density, which is now given by the saturation value
(\ref{dmmax}), is too low to account for (all) dark matter. 

The lowest possible gravitino mass in this scenario is therefore obtained by
setting $T_{\rm R}$ to its lower bound of 200 GeV and saturating the
constraint (\ref{en2}), which gives
\beq \label{en3}
m_{3/2} > 7 \ {\rm keV} \cdot \left( \frac{f_R} {0.5} \right)^{3/7} \, .
\eeq
It should be noted, however, that gravitino masses below 100 keV lead to
over--production of gravitinos from the decay of visible sector superparticles
unless the reheat temperature is below the mass scale $M_{\rm SUSY}$ of these
superparticles \cite{bbn-gravi}. If we demand that this source of gravitinos
contributes at most 10\% of the total dark matter density, gravitino masses
below 1 MeV would have to be excluded, if $T_{\rm R} \gsim M_{\rm SUSY}$. 
Eq.~(\ref{en1}) then requires $B > 400$ GeV (for $f_R = 0.5$).

The upper limit of the allowed range of gravitino masses is also reached for
$T_{\rm R} = 200$ GeV, but depends strongly on the largest value of $B$ one is
willing to contemplate. For example, taking $B = 20$ TeV, eq.(\ref{en1}) would
permit gravitino masses up to 17 TeV. However, since the gravitino is assumed
to be the LSP, a gravitino mass well above 1 TeV does not seem plausible.

We shall notice that unlike the more conventional LSP candidates (such as 
neutralino) there is no prospect for direct detection~\cite{jung} of 
gravitino dark matter. Indeed, this holds for all dark matter candidates 
which have only gravitational interaction with matter~\cite{gravi-dm}. 
However, indirect detection of DM will in this case be possible through the 
effects of NLSP decay on BBN, CMB and diffuse photon 
background~\cite{gravi-dm}.    

%%%%%%%%%%%%%%%%%%%%%%%%%%%%%%%%%%%%%%%%%%%%%%%%%%%%%%%%%%%%%%%%%%%%%%%%%%%
%%%%%%%%%%%%%%%%%%%%%%%%%%%%%%%%%%%%%%%%%%%%%%%%%%%%%%%%%%%%%%%%%%%%%%%%%%%

\section{Lepton asymmetry from sneutrino decay}
\setcounter{footnote}{0}

In the standard (supersymmetric) leptogenesis scenario, the decay of a RH
(s)neutrino with mass $M_i$ (we choose $M_1 < M_2 < M_3$) results in a lepton
asymmetry per (s)neutrino quanta $\epsilon_i$, given by
\beq \label{asymmetry}
\epsilon_i = - {1 \over 8 \pi} {1 \over [{\bf h} {\bf h}^{\dagger}]_{ii}} 
\sum_{j} {\rm Im}
\left( [{\bf h} {\bf h}^{\dagger}]_{ij}\right)^2 f \left({M^{2}_{j}
\over M^{2}_{i}} \right)\,,
\eeq          
with~\cite{one-loop}
\beq \label{corrections}
f(x) = \sqrt{x} \left ({2 \over x-1} + {\ln} \left[{1 + x \over
x}\right ] \right ).
\eeq
The first and second terms on the right-hand side of eq.~(\ref{corrections})
correspond to the one-loop self-energy and vertex corrections, respectively.
Assuming hierarchical RH (s)neutrinos (so that the asymmetry is produced only
from the decay of the lightest RH states), and an ${\cal O}(1)$ $CP$ violating
phase, it can be shown that~\cite{di}
\beq \label{hierarchical}
\vert \epsilon_1 \vert \lsim {3 \over 8 \pi} {M_1 (m_3 - m_1) \over 
{\langle H_u 
\rangle}^{2}},
\eeq
where $m_1 < m_2 < m_3$ are the masses of light, mostly left-handed (LH)
neutrinos. A hierarchical spectrum for RH (s)neutrinos strongly suggests a
hierarchical spectrum of LH neutrino masses, otherwise a big conspiracy would
be required to obtain the latter. We will then have
\beq \label{epsilon1}
\vert \epsilon_1 \vert \lsim 2 \cdot 10^{-10} \left({M_1 \over 
10^6~{\rm GeV}}\right).
\eeq
To obtain the last result, we have used $m_3 \simeq 0.05$ eV (as suggested by
atmospheric neutrino oscillations) and $\langle H_u \rangle \simeq 170$ GeV.
After taking the conversion by sphalerons into account, we arrive at
\beq \label{baryonstandard}
\left({n_{\rm B} \over s}\right)_{\rm standard} \simeq 1.8 \cdot 10^{-10} 
\left({T_{\rm R} \over 10^6~{\rm GeV}}\right) \, .
\eeq
Obtaining the observed baryon asymmetry from the decay of ${\tilde N}_1$
therefore requires
\beq \label{trstandard}    
T_{\rm R} \gsim 10^6~{\rm GeV}
\eeq
even if the (s)neutrinos dominate the energy density of the
universe.\footnote{In thermal leptogenesis this condition is not satisfied;
hence much larger values of $M_1$ and $T_{\rm R}$ are needed in that
scenario \cite{plumacher,buchmuller}.} 

We saw in the previous section that such a high reheat temperature is only
marginally compatible with sufficient gravitino production from sneutrino
decay. Thus, we should consider other leptogenesis mechanisms which generate a
sufficient asymmetry at a lower $T_{\rm R}$. One possibility is to have nearly
degenerate (s)neutrinos such that $\vert M_2 - M_1 \vert \ll M_1$.  In this
case, the $s-$channel pole in the self-energy correction enhances the
generated asymmetry by a factor of $M_1/\vert \Delta M
\vert$~\cite{pilaftsis}.\footnote{Of course, one expects zero asymmetry for
  exactly degenerate (s)neutrinos. The perturbative results
  in~(\ref{hierarchical}) then breaks down, and exact calculations show that
  no asymmetry will be created~\cite{pilaftsis}.} Then we can afford to reduce
$T_{\rm R}$ by a factor of $\vert \Delta M \vert/M_1$. It is also possible to 
enhance the produciton asymmetry~(\ref{epsilon1}), thus lower $T_{\rm R}$, by 
using a specific neuturino mass model~\cite{strumia}. In this case 
heavier RH neutrinos $N_2$ and $N_3$ make contributions to neutrino masses 
${\tilde m}_{2,3}$ which are much larger than the neutrino masses $m_{2,3}$, 
but cancel each other. Then, for $M_1/M_{2,3} < 100$ and an appropriate 
pattern of Yukawa couplings, the asymmetry can be sufficiently enhanced to 
allow successfuil leptogensis for $M_1 \sim {\cal O}(\rm TeV)$.                

Another, completely different, venue is to consider scenarios which, unlike
standard leptogenesis, rely on soft supersymmetry breaking as a source of $CP$
violation. In some of these models, an asymmetry between $\tilde N$ and
${\tilde N}^{\dagger}$ is created by the combined action of neutrino sector
$A-$ and $B-$terms~\cite{nir1,agr}, or from the $B-$term alone via the
Affleck-Dine mechanism~\cite{ad3}. This asymmetry is then transferred to the
light (s)leptons through $\tilde N$ decays as a result of finite temperature
supersymmetry breaking. These models can generate a sufficient asymmetry for a
rather wide range of $M_N$, including $M_N \sim 1$ TeV~\cite{chun}, but
typically need a suppressed $B-$term $B \ll 1$ TeV. There are also models
where the lepton asymmetry is generated in $\tilde N$ decay, mainly due to
supersymmetry breaking contributions to the vertex
correction~\cite{goran,nir2}.\footnote{All of these models work well with only
  one generation of (s)neutrinos. This is a marked difference from the
  standard leptogenesis where mixing among different generations is a
  necessary ingredient.} A very interesting observation has been made recently
in~\cite{nir2}. There it is shown that the $A-$term and gaugino mass
contributions to the vertex correction can lead to successful low-scale
leptogenesis, while all soft parameters (including $B$) assume their natural
value ${\cal O}(\rm TeV)$. More precisely, if the phase mismatch between $A$
and the electroweak gaugino mass $M_{\widetilde W}$ is ${\cal O}(1)$, there is
a contribution to the asymmetry parameter of order~\cite{nir2}
\beq \label{epsilonsoft}
{\vert \epsilon \vert}_{\rm soft} \sim \alpha_W {\vert A M_{\widetilde W}
 \vert \over M^2_1},
\eeq
where $\alpha_W \sim 0.03$ is the weak [$SU(2)$] coupling constant. This
implies that
\beq 
\label{baryonsoft} \left({n_{\rm B} \over s}\right)_{\rm soft} \lsim
\alpha_W {\vert A M_{\widetilde W} \vert T_{\rm R} \over M^3_N}.  
\eeq
Let us choose typical values $\vert A \vert \simeq |M_{\widetilde W}| \simeq
300$ GeV. Then, even if $T_{\rm R}$ is near its lower bound of $200$ GeV, this
mechanism generates a sufficient asymmetry for $M_N \lsim 2 \times 10^5$ GeV.
A heavier sneutrino will be acceptable for higher reheat temperatures.

Notice that for $M_N \sim {\cal O}(\rm TeV)$, soft leptogenesis mechanisms
that exploit a $B-$term $\ll 1$ TeV~\cite{nir1,agr,ad3,chun} can work as well
as a source of gravitino dark matter. The reason is that the soft mass $m_0$
can in this case result in a sufficiently large mass difference
between the sneutrino and neutrino. This case can be treated by replacing $B
\longrightarrow m_0^2 / M_N$ in Eq.(\ref{dm}).

Finally, as noted earlier, we need $T_{\rm R} > 200$ GeV, so that sphalerons
are still active at the time of $\tilde N$ decay. Since $\Gamma_N = h^2 M_N/(4
\pi) \simeq 4 T_{\rm R}^2/M_P$, this implies for $h^2 \equiv \left[ {\bf h}
  {\bf h}^\dagger \right]_{11}$:
\beq \label{hmin}
h^2 > 8 \cdot 10^{-18} \, \frac {10^5 \ {\rm GeV}} {M_N} \, .
\eeq
On the other hand, a sufficient amount of gravitino dark matter could only be
generated from sneutrino decay if $T_{\rm R} \lsim 10^6$ GeV, which implies
the upper bound (note that $T_{\rm R} < M_N$ for perturbative $\tilde N$
decay) 
\beq \label{hmax1}
h^2 < 2 \cdot 10^{-11} \, \frac {10^6 \ {\rm GeV}} {M_N} \, .
\eeq

To summarize, soft leptogenesis is the most efficient mechanism for
generating a sufficient baryon asymmetry along with gravitino dark matter from
the sneutrino decay. It can work for $M_N \lsim 10^6$ GeV. We will later
discuss what this, together with the bounds (\ref{hmin}) and (\ref{hmax1}),
implies for the masses of light neutrinos. We now turn to the problem of
generating cosmological density perturbations from the sneutrino condensate.

%%%%%%%%%%%%%%%%%%%%%%%%%%%%%%%%%%%%%%%%%%%%%%%%%%%%%%%%%%%%%%%%%%%%%%%%%
%%%%%%%%%%%%%%%%%%%%%%%%%%%%%%%%%%%%%%%%%%%%%%%%%%%%%%%%%%%%%%%%%%%%%%%%%

\section{Density perturbations from sneutrino decay}
\setcounter{footnote}{0}

The detection of CMB anisotropies indicates the presence of coherence over
superhorizon scales, which is a strong indication of an early inflationary
stage~\cite{infl}. To date measurements are in agreement with the simplest
prediction from inflation which is a nearly scale-invariant spectrum of
Gaussian and adiabatic primordial perturbations~\cite{wmap}. Traditionally, it
has been considered that quantum fluctuations of the inflaton field (which are
exponentially stretched during inflation) are responsible for density
perturbations~\cite{fmb}. Then, obtaining perturbations of the correct size 
(about $1$ part in $10^{5}$) will constrain parameters of the inflation 
sector. For example, in the chaotic inflation model with $V(\phi) = 
m^2_{\phi} \phi^2 /2$ this leads to $m_{\phi} \simeq 10^{13}$ GeV.

Models where the inflaton does not generate sufficient perturbations can be
rescued, provided that another scalar field is responsible for this. One
possibility is that a late-decaying scalar field which acquires inflationary
fluctuations dominates the energy density of the universe. Isocurvature
fluctuations of this field, called curvaton, will in this case be converted to
curvature perturbations~\cite{lyth,mt}. 

Another possibility is that the inflaton coupling to matter is controlled by
the VEV of some scalar field which acquires fluctuations during inflation. The
equation of state then changes at slightly different moments in different
parts of the universe at the time of inflaton decay. As a result, this
inhomogeneous reheating will also produce curvature
perturbations~\cite{dgz,lev,emp,a}. Note that inflationary fluctuations of a
scalar field are $\sim H_I$ (recall that $H_I$ is the expansion rate during
inflation). Obtaining acceptable perturbations within the curvaton and
inhomogeneous reheating mechanisms requires that the scalar field VEV be $\sim
10^5 H_I$.

A RH sneutrino can give rise to density perturbations through either of these
mechanisms. The sneutrino can play the role of inflaton in a chaotic inflation
model~\cite{sninfl1,sninfl2}, in which case $M_N \simeq 10^{13}$ GeV will be
needed.  Alternatively, for sufficiently small neutrino Yukawas, $\tilde N$
can dominate the thermal bath from reheating before it decays, thus playing
the role of curvaton~\cite{hmy,sncurv} (for an alternative proposal of a
sneutrino as curvaton, see~\cite{abdel}). The inhomogeneous reheating
mechanism can also be realized if the inflaton has non-renormalizable
couplings to matter fields via $\tilde N$, induced by new
physics~\cite{a,anupam,robert}.\footnote{Indeed gravitational effects are
  expected to generate such couplings suppressed by powers of $M_{\rm P}$, see
  Sec.~V~C.} An additional advantage when sneutrino is the inflaton or
curvaton is that baryon asymmetry of the universe can be directly generated at
reheating.

As we saw in the previous sections, simultaneous production of gravitino dark
matter and baryon asymmetry from sneutrino decay requires a dominating $\tilde
N_1$ with mass $M_N \ll 10^{13}$ GeV, which reheats the universe to $T_{\rm R}
\leq 10^6$ GeV. Sneutrino dominance indicates that $\tilde N$ is either the
inflaton or curvaton.\footnote{A thermally produced $\tilde N$ could dominate
the universe only if it can be produced through interactions that do not
contribute to its decay.} However, the condition $M_N \ll 10^{13}$ GeV rules
out $\tilde N_1$ as the inflaton (unless neutrino Yukawas have sufficient
fluctuations to allow a successful inhomogeneous reheating).

Now let us find the condition for sneutrino dominance. We start from a $\tilde
N$ condensate produced during inflation; let $\tilde N_0$ be the absolute
value of the sneutrino field at the end of inflation. This value, and hence
the energy density stored in the sneutrino field, will remain essentially
constant ($\rho_{\tilde N} \simeq M_N^2 N_0^2$) so long as $H > M_N$. At $H =
M_N$ the ratio of the total energy density of the universe ($= 3 H^2 M_{\rm
  P}^2$) and the energy density in the sneutrino field is therefore $N_0^2/(3
M^2_{\rm P})$. At that time the sneutrino could therefore only dominate the
energy density if $N_0 > M_{\rm P}$, which is difficult to achieve in the
context of supergravity models.  Here we will focus on scenarios with $N_0 <
M_{\rm P}$.

For Hubble parameter $H < M_N$, the energy density of the sneutrinos and their
decay products will be redshifted $\propto R^{-3}$ and $\propto R^{-4}$,
respectively, where $R$ is the scale factor of the universe. The same is true
for the inflatons and their decay products, which, as we just saw, dominate at
$H = M_N$. The sneutrino can therefore only become dominant if the inflatons
decay before the sneutrinos do, i.e. if the inflaton decay width $\Gamma_\phi
> \Gamma_N$. The universe will then go through a radiation dominated epoch,
characterized by the temperature $T_I \simeq \sqrt{\Gamma_\phi M_{\rm P} } /
2$ at $H \simeq \Gamma_\phi$ when inflaton decays are completed.

Let us first discuss the case $\Gamma_\phi < M_N$, which seems more plausible
for a weakly coupled inflaton. This means that sneutrino oscillations start
when the universe is still dominated by inflaton matter. While $M_N > H >
\Gamma_\phi$ the amplitude of oscillations will decrease like $H$; in
the radiation dominated epoch, for $H < \Gamma_\phi$, this will change to a
behavior $\propto H^{3/4}$, i.e. $\rho_{\tilde N} \propto H^{3/2}$. In this
radiation dominated epoch the density of the thermal bath (from the inflaton
decay products) will decrease like $H^2$. The two densities will become
equal at temperature
\beq \label{teq}
T_{\rm eq} = T_{\rm I} {N^2_0 \over 3 M^2_{\rm P}}\, .
\eeq
So far we have ignored $\tilde N$ decays. Clearly sneutrino dominance can only
occur if $\tilde N$ has not yet decayed by the time the inflaton decay
products have cooled down to $T = T_{\rm eq}$. This requires $T_{\rm eq} >
T_{\rm R}$; recall that $T_{\rm R}$ is the temperature after sneutrino decay
under the assumptions that sneutrinos do indeed dominate. This condition reads
\beq \label{dom}    
\left({N_0 \over M_{\rm P}}\right)^2 \geq {3 T_{\rm R} \over T_{\rm I}}.
\eeq

A potentially important issue is gravitino production from the thermal bath
after reheating. Eq.(\ref{lspgrav}) shows that for values of $T_{\rm R}$ and
$m_{3/2}$ which allow successful leptogenesis and gravitino dark matter from
sneutrino decay, thermal production of gravitinos at $T_{\rm R}$ will be
subdominant. It, however, can be significant at the first stage of reheating
when $T = T_{\rm I}$. Gravitino production at this epoch can still be
estimated from eq.(\ref{lspgrav}), if we replace $T_{\rm R}$ by $T_I$ and
multiply with the entropy dilution factor from sneutrino decay, which is given
by $T_{\rm R} / T_{\rm eq}$. We require that these thermally produced
gravitinos amount to at most 10\% of the total dark matter density,
i.e. $\Omega_{3/2}^{\rm thermal} h^2 < 0.01$. This leads to the constraint
\beq \label{dil1}
\left({N_0 \over M_{\rm P}}\right)^2 \geq 0.015 \, \left( \frac {M_{\tilde g}}
  {1 \ {\rm TeV}} \right)^2 \, \frac {10 \ {\rm MeV}} {m_{3/2}} \, \frac
{T_{\rm R}} {200 \ {\rm GeV}} \, .
\eeq
The dilution condition (\ref{dil1}) is stronger than the one for domination 
unless there is no gravitino overproduction at $T_{\rm I}$. Finally, we can
use eq.(\ref{en1}) to express the gravitino mass in terms of $B$ and 
$T_{\rm R}$, to find:
\beq \label{dil2}
{N_0 \over M_{\rm P}} \geq 0.02  \, \frac  {M_{\tilde g}} {1 \ {\rm TeV}} \, 
\frac {T_{\rm R}} {200 \ {\rm GeV}} \, \sqrt{\frac {0.5}{f_R}} \, 
\left( \frac {1 \ {\rm TeV}} {B} \right)^2 \, .
\eeq
Requiring $N_0 < M_{\rm P}$ and $M_{\tilde g} > 250$ GeV (from collider
searches \cite{glubound}) then implies $B \gsim 120$ GeV for our canonical
choice $f_R = 0.5$, very similar to the lower bound found in (\ref{en2}). If
the RH neutrinos are charged under some extension of the SM gauge groups, as
e.g. in an $SO(10)$ model, values of $N_0$ near the scale of Grand
Unification, $M_{\rm GUT} \simeq 2 \times 10^{16}$ GeV, seem more natural;
this would require $B \gsim 2$ TeV. Finally, for $B = 20$ TeV, the constraint
(\ref{dil2}) allows values of $N_0$ as small as $10^{14}$ GeV, if $T_{\rm R}$
is near 200 GeV. Recall, however, that the combination of large $B$ and small
$T_{\rm R}$ requires quite large gravitino masses, see eq.(\ref{en1}). Since
heavy gravitinos have a relatively low thermal relic density, the true bound
on $N_0$ is then often set by the condition (\ref{dom}); in particular, $N_0 =
10^{14}$ GeV is only possible for $T_I \geq 3 \cdot 10^{11}$ GeV.
  
Generating density perturbations of the correct size via the curvaton
mechanism requires~\cite{lyth,mt}
\beq \label{pert}
N_0 \sim 10^5 H_I.
\eeq
For values of $N_0$ which satisfy~(\ref{dil2}), this gives $H_I \geq 10^9$
GeV, and we typically have $H_I \simeq 10^{11}-10^{12}$ GeV. Note that in a
chaotic inflation model with such scales inflaton fluctuations do not give
rise to sufficient perturbations, and hence another source will indeed be
needed.

So far we have assumed that $\tilde N$ starts oscillating during the matter
dominated epoch. If the first stage of reheating completes when $H > M_N$, the
expressions in~(\ref{teq}),~(\ref{dom}) and~(\ref{dil1}) will accordingly
change. Then, $T_{\rm I}$ in~(\ref{teq}) and~(\ref{dom}) should be replaced by
the temperature at the onset of sneutrino oscillations $T_{\rm osc} \simeq 0.5
(M_N M_{\rm P})^{1/2}$. Also, the right-hand side of Eq.~(\ref{dil1}) should
be multiplied by a factor of $T_{\rm I}/T_{\rm osc}$.  The largest lower bound
on $N_0$ is in this case obtained for an instant inflaton decay, where $T_{\rm
  I} \simeq 0.5 (H_I M_{\rm P})^{1/2}$. After using $H_I \sim 10^{-5} N_0$ and
$M_N < 10^5$ GeV so that Eq.(\ref{baryonsoft}) generates a sufficient amount
of baryon asymmetry, it is seen that $N_0 < M_{\rm P} \ (M_{\rm GUT})$
requires $B \gsim 0.5 \ (2.5)$ TeV. However, smaller values of $B$ are again
possible if $T_I$ does not saturate its upper bound.

To summarize, successful sneutrino domination which dilutes thermally produced
gravitinos from inflaton decay, along with generating density acceptable
perturbations, can be achieved for $N_0 \geq 10^{14}$ GeV and $H_I \geq 10^9$
GeV.

One comment is in order before closing this section. In our scenario 
sneutrino decay leads to generation of adiabatic baryon and dark matter 
perturbations. Any amount of baryon asymmetry and dark matter created before 
sneutrino decay (for example, via thermal leptogenesis and thermal gravitino 
production) will contribute as an isocurvature component. For a 
sufficiently late-decaying sneutrino condensate, as considered above, this 
isocurvature component will be well below the current bounds from 
WMAP~\cite{iso}. Future CMB experiments, like PLANCK, will provide a tighter 
bound on (and might even detect) such a subdominant component.     
           
%%%%%%%%%%%%%%%%%%%%%%%%%%%%%%%%%%%%%%%%%%%%%%%%%%%%%%%%%%%%%%%%%%%%%%%%%%%
%%%%%%%%%%%%%%%%%%%%%%%%%%%%%%%%%%%%%%%%%%%%%%%%%%%%%%%%%%%%%%%%%%%%%%%%%%%

\section{Additional considerations}
\setcounter{footnote}{0}
\subsection{Thermal effects}

Since inflaton decay can result in a very high reheat temperature $T_{\rm I}$
at the end of the first stage of reheating, one has to be careful about
thermal effects on the dynamics of the sneutrino condensate. During reheating,
the thermal bath consisting of thermalized inflaton decay products has an even
higher temperature $T \simeq 0.7 (H T^2_{\rm I} M_{\rm P})^{1/4} >
T_{\rm I}$~\cite{kt}. During this (inflaton matter dominated) epoch, the
temperature therefore scales like $t^{-1/4}$, i.e. $H \propto T^4$, while $H
\propto T^2$ after the first stage of reheating is completed.

In our case the only thermal effect one has to worry about is early
oscillations of the condensate induced by a large thermal $\tilde N$
mass~\cite{thermal}, which would dilute the sneutrino density before it
decays. If $h N_0 < T$, the Higgs/Higgsino and LH (s)leptons will be in
thermal equilibrium before the onset of sneutrino oscillations. This, in turn,
induces a thermal mass $h T$ for $\tilde N$ which will lead to its early
oscillations, if $h T > H$ for some $H > M_N$. Since $H$ grows faster than
linearly with $T$, such thermal oscillations can be avoided if $h T < H$ when
$H \simeq M_N$. Thermal effects are most severe if the universe is
radiation-dominated at this moment, i.e. if the inflation decay width
satisfies $\Gamma_\phi > M_N$, in which case $H(T) \simeq 5 T^2 / M_{\rm P}$.
The sufficient condition to prevent early oscillations then reads
\beq \label{noearly}
h^2 \leq 2 \cdot 10^{-13} {M_N \over 10^5 \ {\rm GeV}} \, ,
\eeq
which is somewhat stronger than the constraint (\ref{hmax1}). However, the
limit (\ref{noearly}) can be evaded if $h N_0 > T_{\rm max}$. Moreover, it
becomes weaker by a factor $2 \sqrt{M_{\rm P} M_N} / T_I$ in the perhaps more
realistic case $\Gamma_\phi < M_N$, which implies $T_I < \sqrt{M_{\rm P}
M_N}/2$.

We note that the upper bound (\ref{noearly}) implies that scattering between
thermal quanta and the $\tilde N$ condensate \cite{ad3} are not in
equilibrium, i.e. the sneutrino field will indeed oscillate coherently until
it decays. In any case, the soft leptogenesis mechanism of \cite{nir2} works
just as well for an incoherent ensemble of RH sneutrinos.

Finally, we should mention the possibility of non-perturbative ``preheating''
decay of the sneutrino condensate~\cite{ad3,pm}, which can rapidly transfer
energy density from the $\tilde N$ condensate to a plasma (including $\tilde
N$ quanta) with energy up to $\sqrt{h N_0 M_N}$, which might exceed $M_N$
significantly ~\cite{preheat1,preheat2}. It has the undesirable consequence 
that the
sneutrino energy density in this case redshifts faster than that of a
coherently oscillating condensate; moreover, since this mechanism prefers
bosonic final states, if would suppress gravitino production. As noted
in~\cite{ad3}, thermal effects can kinematically block preheating, provided
that $h N_0 M_N< T^2$ at the onset of $\tilde N$ oscillations. Again this
constraint is more stringent for $\Gamma_\phi > M_N$, where it implies
\beq \label{nopreheat1}
h < \frac{M_{\rm P}} {5 N_0} \, .
\eeq 
This bound is much weaker than the one in (\ref{hmax1}) for all $N_0 \lsim
M_{\rm P}$.

%%%%%%%%%%%%%%%%%%%%%%%%%%%%%%%%%%%%%%%%%%%%%%%%%%%%%%%%%%%%%%%%%%%%%%%%%%
\subsection{Initial VEV of the sneutrino}

As shown in Sec.~IV, we need ${\tilde N}$ to have a rather large value $N_0
\gsim 10^{14}$ GeV at the end of inflation. In general the evolution of
$\tilde N$ during inflation is determined by the interplay between quantum
fluctuations in deSitter spacetime \cite{infl} and the classical equation of
motion determined by the sneutrino potential. Here supersymmetry breaking
plays an important role, since it generates a contribution $C_I H_I^2 |\tilde
N|^2$ to the potential \cite{drt1}.

A large value of $N_0$ is most easily generated if $C_I < 0$. In this case
$\tilde N$ will quickly settle in the minimum of the potential, which is
determined by $C_I$ as well as the terms that stabilize the potential for
large $|\tilde N|$. For example, if $N$ transforms non-trivially under the GUT
gauge group, we expect $N_0 \sim M_{\rm GUT}$, as mentioned earlier. For $C_I <
0$, $N_0 \sim 10^5 H_I$, as required if $\tilde N$ is to serve as curvaton,
can be achieved even in ``minimal'' inflation, which only lasts some 60
e-folds.

On the other hand, $C_I > 1$ would imply $N_0 \lsim H_I$, which is not
sufficient for our purpose. Finally, if $|C_I| \ll 1$, the $\tilde N$ mass
during inflation will still be given by $M_N$. For a sufficiently long period
of inflation, $N_0 \simeq H_I^2/5 M_N$ will then be obtained \cite{infl} at
the end of this epoch, regardless of the initial value of $\tilde N$. In this
case $N_0 \sim 10^5 H_I$ implies $M_N \simeq 2 \cdot 10^{-6} H_I \simeq 2
\cdot 10^{-11} N_0$. Recall from the discussion of Sec.~III that soft
leptogenesis requires $M_N \lsim 10^6$ GeV, which implies $N_0 < 2 \cdot
10^{17}$ GeV in this scenario, comfortably in the range of values we found in
Sec.~IV. However, additional positive terms in the potential could reduce
$N_0$. The value $2 \cdot 10^{17}$ GeV is therefore an upper bound for $N_0$
if $C_I \geq 0$.

%%%%%%%%%%%%%%%%%%%%%%%%%%%%%%%%%%%%%%%%%%%%%%%%%%%%%%%%%%%%%%%%%%%%%%%%%%
\subsection{Sneutrino VEV and inflaton decay}

Sneutrino domination for $N_0 \ll M_{\rm P}$ requires that the inflatons decay
rather early. This corresponds to a high reheat temperature $T_{\rm I}$ for
the first stage of reheating. In particular, $T_{\rm I} \geq 10^{10}$ GeV, if
the inflaton decays before $\tilde N$ starts oscillating (assuming that $M_N
\geq 1$ TeV). Obtaining such high reheat temperatures requires that some
inflaton couplings be sufficiently large. Couplings of the inflaton to other
fields are model-dependent and vary from case to case. Recently, it has been
noticed that efficient reheating does not necessarily require large
fundamental couplings in the Lagrangian~\cite{robert}. A scalar condensate
with a sufficiently large VEV can result in a large effective coupling of the
inflaton, even if fundamental couplings are $M_{\rm P}$ suppressed. Here, we
focus on the realization of this scenario, called enhanced reheating, in the
presence of the sneutrino condensate. Indeed it will be interesting that
$\tilde N$ can set the stage for its domination by facilitating inflaton
decay.

To elucidate, we consider a simple superpotential term
\beq \label{enhsup}
W \supset \lambda {\Phi \over M_{\rm P}} {\bf N} {\bf H}_u {\bf L},
\eeq
where $\Phi$ is the inflaton superfield. Regardless of any discrete or
continuous global symmetry, gravitational effects are expected to generate
such a term in the superpotential. If $\Phi$ remains a gauge singlet up to
$M_{\rm P}$, $\lambda \sim {\cal O}(1)$ is expected. On the other hand, if the
inflaton is non-singlet under some extended gauge group which is broken at a
scale $M_{\rm new}$, $\lambda$ will be suppressed by powers of $M_{\rm
  new}/M_{\rm P}$.

The rate for inflaton decay from~(\ref{enhsup}) reads
\beq \label{enhdec}
\Gamma_{\phi} \simeq {\lambda^2 \over 4 \pi} \left({N_0 \over 
M_{\rm P}}\right)^2 m_{\phi}.
\eeq
For chaotic inflation, $m_{\phi} = H_I$, and the condition for acceptable 
perturbations via curvaton mechanism $N_0 \sim 10^5 H_I$ leads to
\beq \label{enhreh}
T_{\rm I} \simeq 4 \times 10^{-4} \lambda {N^{3/2}_0 \over \sqrt{M_{\rm P}} }.
\eeq
If the inflaton is an absolute singlet, $\lambda \simeq 1$. For $N_0 \simeq
M_{\rm GUT} \simeq 2 \cdot 10^{16}$ GeV, we will then have $T_{\rm I} \sim 7
\cdot 10^{11}$ GeV. Therefore, a sneutrino VEV which is typically preferred
by our scenario, can naturally lead to a fast inflaton decay as needed in the
curvaton mechanism.
                     
%%%%%%%%%%%%%%%%%%%%%%%%%%%%%%%%%%%%%%%%%%%%%%%%%%%%%%%%%%%%%%%%%%%%%%%%%%
\subsection{Non-dominating sneutrino}

So far our results have been derived for the case when the sneutrino
condensate dominates the universe before decaying. This is necessary if the
sneutrino is to serve as curvaton. In this section we investigate if RH
sneutrinos that do {\em not} dominate the energy density of the universe can
simultaneously produce leptogenesis and gravitino dark matter. Since there
will be only one stage of reheating, it will be more appropriate to replace
$T_{\rm R}$ and $T_{\rm I}$ with the sneutrino decay temperature\footnote{Note
that this is now simply the temperature at which the sneutrino decays, i.e.
the temperature of the thermal plasma at $H = \Gamma_N$. By assumption this
temperature is not changed significantly by these decays. We still assume
that the sneutrino decays after the inflaton does.}  $T_{\rm d}$ and the
universe reheat temperature $T_{\rm R}$, respectively.  Note that $T_{\rm R}$
should now be sufficiently low so that thermal gravitino
production~(\ref{lspgrav}) is subdominant.  

The sneutrino energy density at $H \simeq \Gamma_N$ will now be a fraction $r
< 1$ of the energy density of the thermal bath from inflaton decay. For
$\Gamma_\phi < M_N$\footnote{Recall that early inflaton decays more easily
lead to sneutrino domination; hence we consider relatively late inflaton
decays here.} this fraction can be computed from
\beq \label{r}
r = {T_{\rm R} N^2_0 \over 3 T_{\rm d} M^2_{\rm P}}.
\eeq
The number density of gravitinos produced in sneutrino decay is reduced from
(\ref{dm}) to (we take $f_R = 0.5$ from now on)
\beq \label{dm2}
\left({n_{3/2} \over s}\right)_{\rm decay} \simeq 8 \cdot 10^{-3} r {B^4 \over
  m^2_{3/2} T_{\rm d} M_{\rm P}}\, .
\eeq
The upper bound on the gravitino number density from phase space is still
given by Eq.~(\ref{dmmax}), with $T_{\rm R} \rightarrow T_{\rm d}$. The
gravitino density from sneutrino decay will saturate this bound for gravitino
mass 
\beq \label{satur2}
m^{\rm satur}_{3/2} = 5 \left({r B T^2_{\rm d} \over M_{\rm P}} \right)^{1/2}.
\eeq

Eq.~(\ref{dm2}) gives the correct dark matter density for
\beq \label{en4}
m_{3/2} = 56 \, {\rm MeV} \, r \, \left( \frac {B} {1 \ {\rm TeV}} \right)^4 \,
\frac {200 \ {\rm GeV}} {T_{\rm d}} \, .
\eeq
This will be above the saturation value (\ref{satur2}) if
\beq \label{rmin}
r \geq 1.4 \cdot 10^{-7} \, \left( \frac {1 \ {\rm TeV}} {B} \right)^7 \,
\left( \frac {T_{\rm d}} {200 \ {\rm GeV}} \right)^4 \, .
\eeq
The lowest possible value of the gravitino mass compatible with gravitino dark
matter from sneutrino decay is obtained when the bound (\ref{rmin}) is
saturated:
\beq \label{mgravmin}
m_{3/2} \geq 8 \ {\rm eV} \, \left( \frac {1 \ {\rm TeV}} {B} \right)^3
 \, \left( \frac {T_{\rm d}} {200 \ {\rm GeV}} \right)^3 \, .
\eeq
The constraint (\ref{rmin}) would allow values of $r$ below $10^{-14}$ for $B
> 10$ TeV, if $m_{3/2}$ is well below 1 eV. However, as already noted at
the end of Sec.~II, gravitinos of such low mass would be over-produced from
the decays of visible sector sparticles \cite{bbn-gravi}. Requiring $m_{3/2} >
1$ MeV in order to keep this contribution subdominant even if sparticles are
light, and again taking $B \lsim 20$ TeV from naturalness argument, we see
from (\ref{en4}) that $r \gsim 10^{-7}$ is required.

We now turn to sneutrinos as source of the baryon asymmetry. We saw in
Sec.~III that standard high scale leptogenesis, where $CP$ violation comes from
the neutrino Yukawa couplings, is only marginally compatible with gravitino DM
from sneutrino decay even if sneutrinos dominate the universe, unless two (or
more) RH sneutrinos are closely degenerate, or for a specific neutrino mass 
model. Here we therefore focus on soft
leptogenesis. Eq.~(\ref{baryonsoft}) is modified to
\beq \label{baryonsoft2}
\left({n_{\rm B} \over s}\right)_{\rm soft} \lsim \alpha_W r {\vert A 
M_{\widetilde W} \vert T_{\rm d} \over M^3_N}\, .
\eeq
Note that this expression is only valid for $M_N^2 > \vert A M_{\widetilde W}
\vert $. Since $\alpha_W \simeq 0.03$, leptogenesis therefore also requires $r
\gsim 10^{-7}$. This bound can be saturated only for values of $M_N$ not much
in excess of 1 TeV. In that case the term $\propto m_0^2$ in the scalar
potential (\ref{low}) can also contribute significantly to the
sneutrino-neutrino mass splitting. The desired gravitino density can then be
obtained for smaller values of $B$.

Finally, the curvaton mechanism will not work if $\tilde N$ does not dominate.
The sneutrino condensate can nevertheless play a crucial role in the
generation of density perturbations via inhomogeneous reheating proceeding
through the coupling in~(\ref{enhsup}). It is interesting that for a smaller
$N_0$, usually needed to have $r < 1$, this coupling also results in a lower
$T_{\rm R}$ required to avoid thermal overproduction of gravitinos. Of course,
we still need $N_0 \simeq 10^5 H_I$ for this mechanism to work. Scenarios with
$r \ll 1$, and correspondingly small values of $N_0$, would therefore require
a relatively low Hubble parameter during inflation. For example, taking
$T_{\rm R} \sim 10^3 T_{\rm d}$, eq.(\ref{r}) shows that $H_I \sim 10^8$ GeV
would be required if $r$ is near its lower limit of $10^{-7}$.

%%%%%%%%%%%%%%%%%%%%%%%%%%%%%%%%%%%%%%%%%%%%%%%%%%%%%%%%%%%%%%%%%%%%%%%       
\section{Summary and Conclusions}
\setcounter{footnote}{0}

We have shown that an SU(2) singlet ``right-handed'' sneutrino field $\tilde
N$, which has originally been introduced in the context of supersymmetric
see-saw models of neutrino masses, can be the simultaneous source of the
baryon asymmetry (via leptogenesis), of nonthermal dark matter (via its decay
into gravitinos), and of density perturbations (via either the curvaton or
inhomogeneous reheating mechanism).  This can most easily be achieved if the
sneutrinos dominated the total energy density of the universe for some period
of time after inflation. Dark matter production from this source requires a
relatively low reheat temperature after sneutrino decay ($T_{\rm R} \lsim
10^6$ GeV). This is only marginally compatible with standard leptogenesis,
where the necessary $CP$ violation comes from the neutrino Yukawa couplings,
but comfortably fits in the range of temperatures where ``soft leptogenesis''
can work, where the source of $CP$ violation is in the soft breaking terms. On
the other hand, our mechanism can work for a wide range of gravitino masses (1
MeV $\lsim m_{3/2} \lsim$ 1 TeV), and only imposes a mild lower bound on the
soft breaking $B$ parameter associated with the large Majorana mass driving
the see-saw mechanism, $B \gsim 400$ GeV.

If the sneutrino is not the curvaton, its energy density need not dominate. In
fact, it can still explain both the baryon asymmetry and gravitino dark matter
if its energy density only amounted to $10^{-7}$ of the total. In this case we
need a lower sneutrino decay temperature, larger $B$, and/or smaller
$m_{3/2}$, compared to the case where the sneutrino dominates.

One perhaps not so attractive aspect of our model (also existing in the model
of Refs.~\cite{hmy,iy}) is that the Yukawa couplings of $\tilde N$ should be
small, see (\ref{hmax1}). Such a small value for $h$ implies that the lightest
LH neutrino $\nu_1$ should be extremely light; for example, taking typical
values $T_{\rm R} = 1$ TeV, $M_N = 100$ TeV we find $m_{\nu_1} \sim 10^{-7}$
eV. Theoretically, it has been suggested that a small Yukawa coupling (as
well as $M_N \ll M_{\rm GUT}$), can be explained in the context of orbifold
GUT models~\cite{hmy1}.  Experimentally, an extremely light (even massless)
$\nu_1$ is allowed and has testable consequences for neutrinoless double beta
decay experiments~\cite{dbd}. Therefore a tiny Yukawa coupling, despite being
not attractive, can be accommodated by theory and experiment.

The sneutrino energy density can easily be large enough for the sneutrino to
serve as curvaton or to trigger inhomogeneous reheating if its Hubble induced
soft SUSY breaking mass during inflation was negative or small. In particular,
we find acceptable scenarios where the sneutrino field after inflation is of
the order of the scale of Grand Unification, as expected e.g. in $SO(10)$
models. If we only want the sneutrinos to produce the baryon asymmetry and
gravitino dark matter, sufficiently many $\tilde N$ quanta could also have
been produced from inflaton decay. However, the sneutrino abundance is
model dependent in this case and can vary considerably. 

It should be admitted that we have not explained why the baryon and dark
matter densities are of comparable magnitude. Even though they have a common
origin, their numerical values depend on different combinations of parameters.
In particular, the dark matter density scales like the fourth power of $B$ and
is inversely proportional to the gravitino mass. These parameters do not
affect the baryon density at all. Conversely, the dark matter density does not
depend on the $CP$ violating phases that crucially enter the expression of the
baryon asymmetry. However, on close inspection this seems to be true for all
models that have been proposed so far. We find it encouraging that for the
soft leptogenesis scenario the main remaining freedom comes from soft breaking
parameters. Within our framework the baryon and dark matter densities can
therefore be considered as cosmological constraints on the mechanism of
supersymmetry breaking. Even in the absence of an explicit model that gives
the required relations between soft breaking parameters we find it encouraging
that we can solve three of the most important problems in current cosmology
using a single field, which has the additional advantage of being well
motivated from particle physics.

%%%%%%%%%%%%%%%%%%%%%%%%%%%%%%%%%%%%%%%%%%%%%%%%%%%%%%%%%%%%%%%%%%%%%%%%%%%
%%%%%%%%%%%%%%%%%%%%%%%%%%%%%%%%%%%%%%%%%%%%%%%%%%%%%%%%%%%%%%%%%%%%%%%%%%%
\section*{Acknowledgements}

The authors are thankful to A. Mazumdar for useful discussions. The work of 
R.A. is supported by the National Sciences and Engineering
Research Council of Canada.

%%%%%%%%%%%%%%%%%%%%%%%%%%%%%%%%%%%%%%%%%%%%%%%%%%%%%%%%%%%%%%%%%%%%%%%%%%
%%%%%%%%%%%%%%%%%%%%%%%%%%%%%%%%%%%%%%%%%%%%%%%%%%%%%%%%%%%%%%%%%%%%%%%%%%

\end{document}